\documentclass[prb,twocolumn,floatfix,superscriptaddress,amsmath,aps,longbibliography]{revtex4-1}
\pdfoutput=1
\usepackage[T1]{fontenc}\usepackage[latin1]{inputenc}
\usepackage{dcolumn,bm,graphicx,xcolor,booktabs,microtype,afterpage,soul} \graphicspath{{figs/}}
\usepackage[charter,greekuppercase=italicized]{mathdesign}
\renewcommand{\figurename}{Fig.}
\renewcommand{\tablename}{Table}
\makeatletter\renewcommand{\fnum@figure}[1]{\figurename~\thefigure.}\makeatother
\makeatletter\renewcommand{\fnum@table}[1]{\tablename~\thetable.}\makeatother
\newcount\hh \newcount\mm
\hh=\time \divide\hh by 60
\mm=\hh \multiply\mm by 60 \mm=-\mm
\advance\mm by \time
\def\now{\number\hh:\ifnum\mm<10{}0\fi\number\mm}
\usepackage[colorlinks,plainpages=false,linkcolor=blue,urlcolor=blue,citecolor=blue,pdfpagemode=UseNone,pdfstartview=FitBH]{hyperref}
\usepackage[version=4]{mhchem}

\begin{document}


\title{Electron-phonon coupling in EuAl$_4$ under hydrostatic pressure}

\author{A.~S.~Sukhanov}
\affiliation{Experimental Physics VI, Center for Electronic Correlations and Magnetism, University of Augsburg, 86159 Augsburg, Germany}

\author{S.~Gebel}
\affiliation{Institut f{\"u}r Festk{\"o}rper- und Materialphysik, Technische Universit{\"a}t Dresden, D-01069 Dresden, Germany}

\author{A.~N.~Korshunov}
\affiliation{Donostia International Physics Center (DIPC), Paseo Manuel de Lardiz\'abal, 20018, San Sebasti\'an, Spain}

\author{N.~D.~Andriushin}
\affiliation{Institut f{\"u}r Festk{\"o}rper- und Materialphysik, Technische Universit{\"a}t Dresden, D-01069 Dresden, Germany}

\author{M.~S.~Pavlovskii}
\affiliation{Kirensky Institute of Physics, Siberian Branch, Russian Academy of Sciences, Krasnoyarsk 660036, Russian Federation}

\author{Y.~Gao}
\affiliation{Department of Physics and Astronomy, Rice University, Houston, Texas 77005, USA}

\author{J.~M.~Moya}
\affiliation{Department of Chemistry, Princeton University, Princeton, New Jersey 08544, USA}

\author{K.~Allen}
\affiliation{Department of Physics and Astronomy, Rice University, Houston, Texas 77005, USA}

\author{E.~Morosan}
\affiliation{Department of Physics and Astronomy, Rice University, Houston, Texas 77005, USA}

\author{M.~C.~Rahn}
\thanks{Corresponding author: marein.rahn@uni-a.de}
\affiliation{Experimental Physics VI, Center for Electronic Correlations and Magnetism, University of Augsburg, 86159 Augsburg, Germany}

\begin{abstract}

In the intermetallic rare-earth tetragonal EuAl$_4$ system, competing itinerant exchange mechanisms lead to a complex magnetic phase diagram, featuring a centrosymmetric skyrmion lattice. Previous inelastic x-ray scattering (IXS) experiments revealed that the incommensurate charge-density wave (CDW) transition in  EuAl$_4$ ($T_{\text{CDW}} = 142$~K) is driven by momentum-dependent electron-phonon coupling (EPC). We present the results of IXS under high hydrostatic pressure induced by diamond anvils and show how the EPC in EuAl$_4$ is renormalized and suppressed in the material's temperature-pressure phase diagram. Our findings highlight the crucial role of momentum-dependent EPC in the formation of the CDW in EuAl$_4$ and provide further insights into how external pressure can be used to tune charge ordering in quantum materials.



\end{abstract}

\maketitle

\section{Introduction}


Charge-density waves (CDWs) have emerged as a significant area of focus in condensed matter research due to their widespread occurrence across various quantum materials. These electronic ordering phenomena are often linked to and can even enhance the properties of other emergent phases, such as those found in transition-metal dichalcogenides~\cite{Ritschel2015}, unconventional superconductors~\cite{Loret:2019aa,Silva-Neto:2014aa, fernandes2022iron}, and systems exhibiting complex magnetic behaviors~\cite{Yasui:2020aa,Teng:2023aa,Chen:2016aa}. The potential of CDWs to coexist and couple with such phases underscores their importance in the study of correlated electron systems.

The Peierls instability serves as the foundational concept for CDW formation in idealized one-dimensional systems. Here, perfect nesting of the Fermi surface allows for the creation of a periodic lattice distortion. This process introduces a CDW gap at the Fermi level, leading to a transition from metallic to insulating behavior. While this theoretical framework is appealing for its simplicity, it assumes conditions rarely found in real-world materials.

In practice, materials frequently diverge from the idealized conditions described by the Peierls scenario. In higher-dimensional systems, such as those with two- or three-dimensional structures, electron-phonon coupling tends to be a more dominant factor than perfect Fermi surface nesting (FSN)~\cite{Johannes:2008aa,Zhu:2017aa}. This is exemplified by materials like NbSe$_2$, which exhibit CDW transitions without accompanying metal-insulator transformations and show no FSN~\cite{Xi:2015aa,Zhu:2015aa}.
In this case, lattice dynamics might play a crucial role in the CDW formation, particularly through the coupling between electronic states and vibrational modes of the crystal lattice. The interaction between these states can lead to phonon softening, a phenomenon characterized by reduced vibrational frequencies at specific wave vectors.

EuAl$_4$, a rare-earth intermetallic, offers a particularly intriguing platform for exploring CDW-driven phenomena. This material undergoes a second-order CDW phase transition at $\sim$142~K, leading to substantial changes in its electronic and structural properties~\cite{Ramakrishnan:gq5015,Shang:2021aa,Shimomura:2018aa,Kaneko:2021aa,Moya:2022aa,PhysRevMaterials.8.104414}. Moreover, EuAl$_4$ hosts a centrosymmetric skyrmion lattice at lower temperatures, showcasing the interplay between charge ordering and magnetism~\cite{Moya:2022aa, Takagi:2022aa, Gen:2023aa,Shimomura:2018aa,Meier:2022aa,Vibhakar2024}. Unlike traditional CDWs that are strongly linked to Fermi surface nesting, the CDW in EuAl$_4$ appears to originate from a distinct mechanism. The previous experimental~\cite{PhysRevB.110.045102} and theoretical~\cite{Wang2024} studies suggest that the driving force is indeed momentum-dependent electron-phonon coupling (EPC). This coupling selectively interacts with specific vibrational modes, leading to the formation of a complex CDW state.

Particularly, inelastic x-ray scattering (IXS) measurements supported by first-principles calculations~\cite{PhysRevB.110.045102} revealed a broad softening of a transverse acoustic (TA) phonon mode along the $\Gamma-Z$ direction, which begins well above room temperature and becomes more pronounced as the system approaches $T_{\text{CDW}}$. Unlike conventional CDW systems where phonon anomalies are sharply localized at a specific wave vector, the softening in EuAl$_4$ was found more extended, indicating that strong momentum-dependent EPC plays a dominant role in driving the transition. First-principles calculations~\cite{PhysRevB.110.045102} identified the atomic displacements in the soft TA mode and showed that the structural modulation observed in the CDW phase are in essence a freezing of these transverse lattice vibrations. 

\begin{figure}[t]
\center{\includegraphics[width=0.85\linewidth]{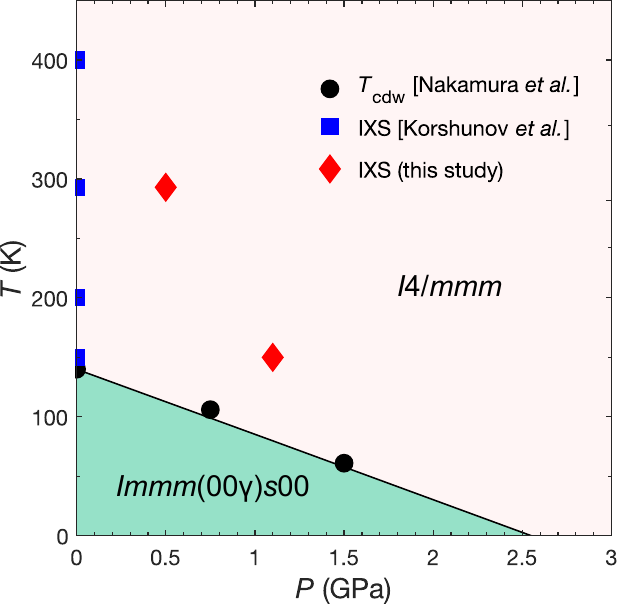}}
  \caption{ The $(T$--$P)$ phase diagram of EuAl$_4$. The CDW transition is shown as reported by Nakamura~\textit{et al.}, Ref.~\cite{Nakamura:2015aa}. The measurements at the $(T$--$P)$ points studied in the present paper are marked by red diamonds.}
  \label{fig:Fig1}
  \vspace{-12pt}

\end{figure}

Nakamura \textit{et al.}~\cite{Nakamura:2015aa} examined the high-pressure evolution of the CDW transition in EuAl$_4$ using resistivity measurements. Electrical resistivity and thermoelectric power measurements revealed that $T_{\text{CDW}}$ decreases progressively with increasing pressure at a rate of d$T_{\text{CDW}}$/d$P = -54.7$\,K/GPa (see Fig.~\ref{fig:Fig1}), eventually vanishing at $\sim$2.5\,GPa. The suppression of the CDW phase suggests that the underlying electron-phonon interactions responsible for its formation are significantly weakened by hydrostatic compression. Besides the suppression of the CDW, no signatures of any other structural transitions were observed~\cite{Nakamura:2015aa}, implying the parent $I4/mmm$ space group remains stable under high pressure.

In this paper, we extend the previous IXS measurements at the ambient pressure to the measurements of the same soft TA phonon mode under high hydrostatic pressure and discuss how the EPC in EuAl$_4$ is renormalized in its ($T$--$P$) phase diagram. Figure~\ref{fig:Fig1} summarizes the temperature and pressure at which the previous~\cite{PhysRevB.110.045102} and the present IXS measurements were carried out.

\section{Experimental details}

High-quality single crystals of \ce{EuAl4} were grown by the self-flux technique, as previously described in Ref.~\cite{Stavinoha:2018aa}. The stoichiometry and crystalline quality were confirmed by energy-dispersive X-ray analysis (EDX) and powder X-ray diffraction. 

Low-energy dispersive lattice fluctuations were characterized by inelastic x-ray scattering at the IXS branch of ID28~\cite{ESRFdataDOI}. The sample was loaded into a diamond anvil cell (DAC) with helium as the pressure-transmitting medium. Pressure was calibrated using ruby luminescence. For low-temperature measurements, the pressure cell was placed inside a helium cryostat.

The spectrometer was operated with a Si~(12 12 12) backscattering monochromator at a wavelength of 0.5226\,\AA ~ (23.7\,keV), which provides an energy resolution of 1.5\,meV in full width at half maximum. IXS energy-transfer scans (at constant momentum transfer) were obtained in transmission geometry for momentum transfers along selected high-symmetry directions of reciprocal space.

All IXS spectra were fitted with a phenomenological model assuming one Stokes and one anti-Stokes excitation per phonon branch, each resolution- function. The instrumental resolution was determined from a spectrum collected in the vicinity of the elastic line of the $(2,0,0)$ Bragg peak. It is well described by a pseudo-Voigt profile with a full width at half-maximum of 1.5~meV. The phenomenological fits, therefore, have three free parameters: the energy of the excitation, its intensity and the lineshape mixing factor. The dataset obtained under high pressure is further compared with the previous data at ambient pressure~\cite{PhysRevB.110.045102}, which was collected in a setup with 3~meV resolution.

\section{Results}

\begin{figure}[t]
\center{\includegraphics[width=1.0\linewidth]{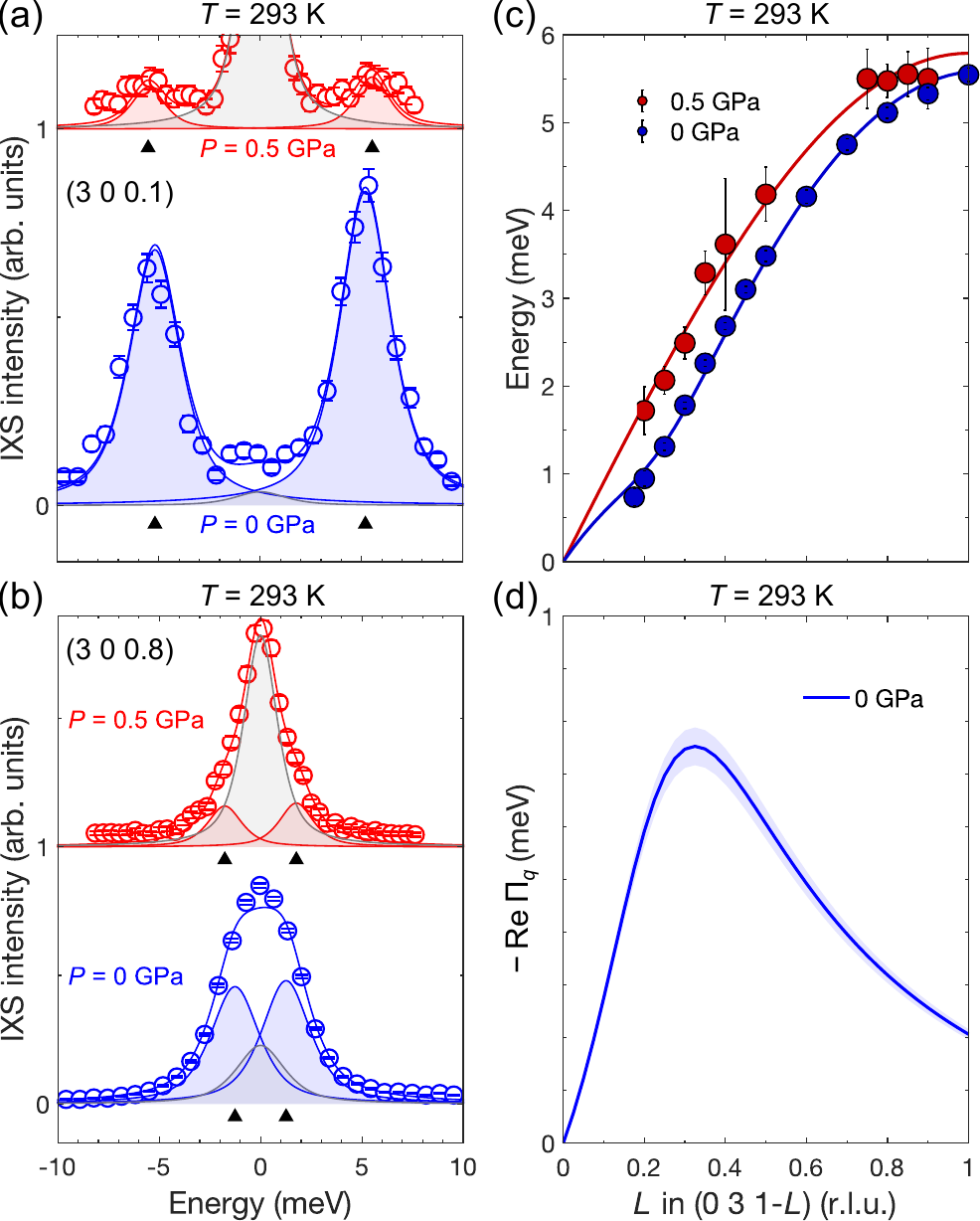}}
  \caption{ Inelastic X-ray scattering at $T = 293$~K. (a) The data collected at the momentum (3~0~0.1) at high pressure of 0.5\,GPa, compared to the previous measurements at ambient pressure~\cite{PhysRevB.110.045102}. (b) The data at (3~0~0.8). The open symbols are data and the solid lines are the pseudo-Voigt fits. The shaded areas illustrate the individual peak contributions into the net fitting curve. The black marks highlight the peak positions. (c) The extracted phonon dispersions at 0 and 0.5\,GPa. The solid symbols are the fit results, and the solid lines describe the model given in Eq.~\ref{eq:eq1}. (d) The real part of the phonon self energy at ambient pressure, according to Eq.~\ref{eq:eq1}, with the shaded band showing the $1\sigma$ margin of uncertainty}
  \label{fig:Fig2}
  \vspace{-12pt}

\end{figure}

\begin{figure}[t]
\center{\includegraphics[width=1.0\linewidth]{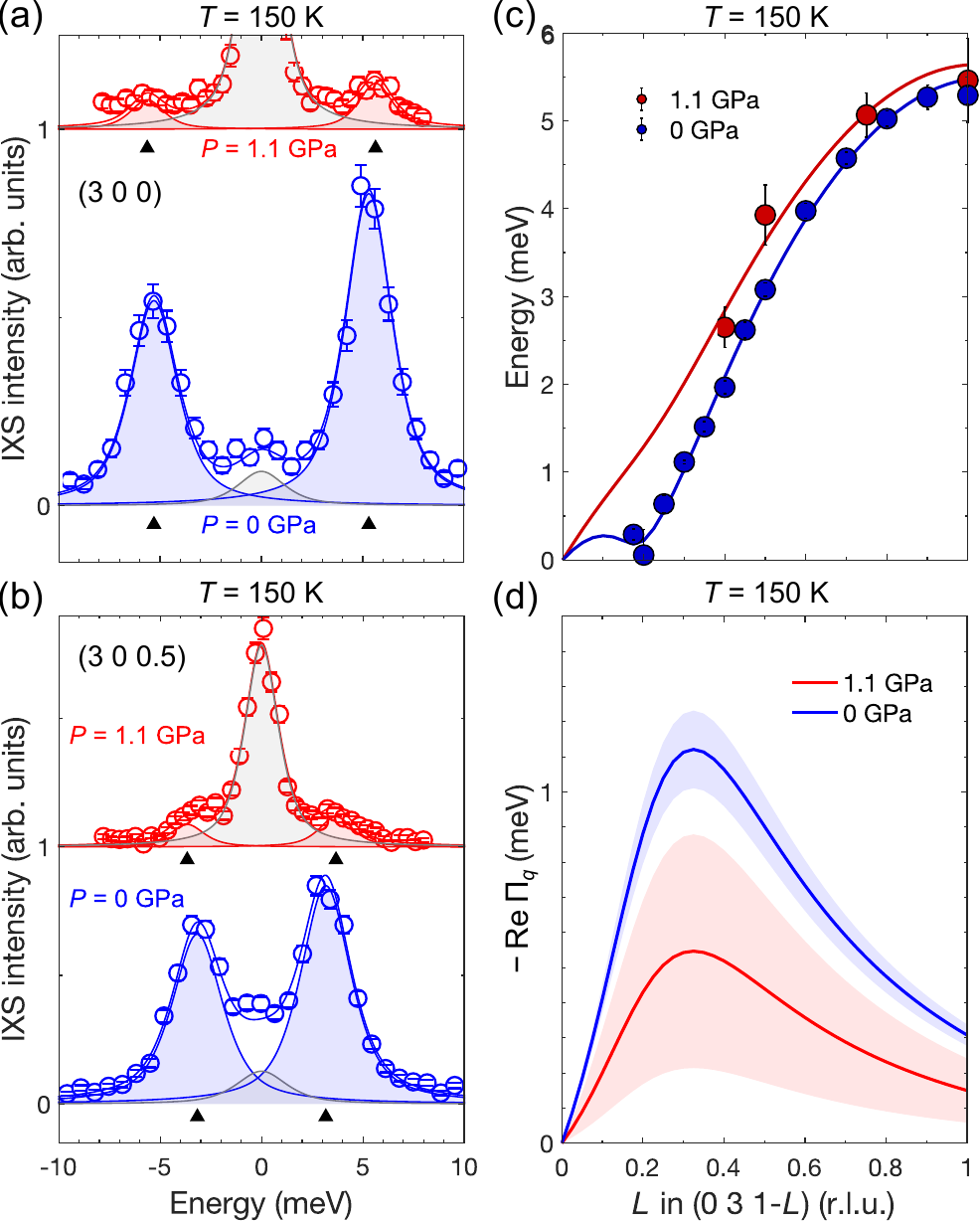}}
  \caption{ Inelastic X-ray scattering at $T = 150$~K. (a) The data collected at the momentum (3~0~0) at 1.1\,GPa, as compared to the previous measurements at  ambient pressure~\cite{PhysRevB.110.045102}. (b) The data at (3~0~0.5). The open symbols are data, the solid lines are the pseudo-Voigt fits. The shaded areas illustrate the individual peak contributions to the net fitting curve. The black markers highlight the peak positions. (c) The extracted phonon dispersions at 0 and 1.1\,GPa. The solid symbols are the fit results, and the solid lines describe the model given in Eq.~\ref{eq:eq1}. (d) The real part of the phonon self energy for 0 and 1.1\,GPa, with shaded bands showing the $1\sigma$ margin of uncertainty.}
  \label{fig:Fig3}
  \vspace{-12pt}

\end{figure}


We begin presenting our experimental results by showing the raw data collected at momenta (3 0 0.1)~r.l.u., which is close to the Brillouin zone boundary of (3 0 0), at room temperature and applied pressure of 0.5\,GPa. In Fig.~\ref{fig:Fig2}(a), the spectrum at high pressure is compared with the spectrum previously collected at the same momentum at ambient pressure~\cite{PhysRevB.110.045102}. Both spectra were fitted by the same model of a pair of the pseudo-Voigt-shaped peaks centered at $\pm \omega$ energy with respect to the central elastic line at $\omega = 0$. The data obtained under pressure contains a larger contamination of the elastic peak, which can be attributed to elongated Bragg tails of a reflection from the diamond anvil. Nevertheless, fitting the model allows an accurate determination of the phonon energy. As can already be seen at a glance, the application of 0.5\,GPa does not induce a significant phonon energy shift at the Brillouin zone boundary. This implies that the trivial pressure-induced mode hardening is almost negligible.

Next, we turn to the comparison of the two spectra at $P = 0$ and 0.5\,GPa, at a momentum transfer corresponding to the position of the CDW propagation vector at $L = 0.2$~r.l.u., see Fig.~\ref{fig:Fig2}(b). At this momentum transfer, the mode softening is maximized, as observed in previous IXS measurements~\cite{PhysRevB.110.045102}, and it is already sizable at room temperature. IXS spectra collected at 0.5\,GPa clearly show that the phonon positions shifted to higher energies, which points to a significant suppression of the EPC already at such moderate pressures.

In order to fully analyze the renormalized dispersion under 0.5~GPa, in Fig.~\ref{fig:Fig2}(c) we plotted the extracted IXS peak positions next to the previously reported ambient pressure dispersion~\cite{PhysRevB.110.045102}. Indeed, the whole phonon branch shows higher energy for all the momenta across the Brillouin zone. The pressure-induced shift in energy is higher around the momenta at which the softening is centered at 0~GPa, and diminishes towards the Brillouin zone edge, where the EPC plays a lesser role.

We note, for reference, that previous theory studies~\cite{Wang2024} simulated the TA mode of BaAl$_4$ along $\Gamma$--$Z$, which featured no EPC, and thus no phonon softening and no CDW transition~\cite{PhysRevB.94.224306,Wang2021,Ramakrishnan2024}. In this case, the simulated dispersion has no down-turn anomalies and can be well described by a simple relation $\omega = a\sin{\left( \pi L/2 \right)}$, where $a$ is the scaling coefficient that defines the mode bandwidth.

The red solid line in Fig.~\ref{fig:Fig2}(c) shows a fit by this sine function, with $a = 5.79$\,meV. Evidently, the dispersion at 0.5\,GPa can be considered as the bare dispersion with no influence of EPC. At ambient pressure, this scenario is not reached even at 400~K ($2.8\times T_{\text{CDW}}$)~\cite{PhysRevB.110.045102}. If we assume that the observed spectrum of 0.5~GPa is close to the bare dispersion $\omega_{0,L} = a\sin{\left( \pi L/2 \right)}$, then the EPC-renomalized dispersion can be modelled by the equation~\cite{PhysRevB.94.224306}:
\begin{equation}
\omega^2_L = \omega_{0,L}^2 + 2\omega_{0,L} \text{Re}\Pi_{L,\omega},
\label{eq:eq1}
\end{equation}
Here, $\text{Re}\Pi_{L,\omega}$ is the real part of the phonon self energy, which generally depends on both the momentum $L$ and energy $\omega$.

To describe the phonon softening at different pressures and temperatures, we model the phonon self energy by the following simplified energy-independent equation:
\begin{equation}
\text{Re}\Pi_{L} = \frac{2A}{\pi} \sin{\left( \pi L/2 \right)}  \frac{w}{4\left(L - L_0\right)^2 + w^2},
\label{eq:eq2}
\end{equation}
where the free parameters $L_0 = 0.18$~r.l.u. and $w = 0.57$~r.l.u. were determined via the global fit of all the dispersion for all $P$ and $T$, whereas the coupling strength $A$ is the parameter fitted for each individual dispersion. The solid blue line in Fig.~\ref{fig:Fig2}(c) for the dispersion at 0.5\,GPa shows that Eq.~\ref{eq:eq1} describes our observations well, with only one free parameter ($A$). Figure~\ref{fig:Fig2}(d) illustrates the extracted function $\text{Re}\Pi$ for the spectrum collected at ambient pressure. The model EPC maximizes the phonon softening at $L \sim 0.3$~r.l.u. and has a broad width, extending through most of the Brillouin zone.

Next, we can apply this model of EPC in EuAl$_4$ to low temperatures, where the phonon softening is expected to be more prominent. Figures~\ref{fig:Fig3}(a) and \ref{fig:Fig3}(b) compare the IXS data collected at 150\,K at the zone-boundary point (300) [Fig.~\ref{fig:Fig3}(a)] and a middle-zone momentum (3~0~0.5) [Fig.~\ref{fig:Fig3}(b)], with and without applied pressure. Measurements in the Brillouin zone center, i.e. at low photon energies, were hindered by overlap with the strong elastic signal of the diamond anvils. The comparison of the mode softening at 150~K at pressures of 0 and 1.1~GPa is therefore limited to (3~0~0.5) as the closest momentum to the CDW propagation vector. 

Nevertheless, the spectra in Fig.~\ref{fig:Fig3}(b) confirm that the phonon softening is suppressed by high pressure as the positions of the IXS peaks shift towards higher energies. Similarly to the high-pressure results obtained at room temperature, the data at 150\,K show negligible change in the phonon energies at the zone boundary [Fig.~\ref{fig:Fig3}(a)]. The fitted phonon energies at different momenta at 1.1\,GPa are summarized in Fig.~\ref{fig:Fig3}(c), where we compare the pressure-renormalized dispersion with the pristine dispersion at ambient pressure. By applying the same model of the EPC given by the Eq.~\ref{eq:eq1}, we extract the real part of the phonon self energy at 150\,K, as shown in Fig.~\ref{fig:Fig3}(d). As can be seen, the application of 1.1\,GPa results in suppression of the EPC in EuAl$_4$ by a factor of $\sim$2.

\section{Lattice-dynamics simulations under pressure}

\begin{figure}[t]
\center{\includegraphics[width=1.0\linewidth]{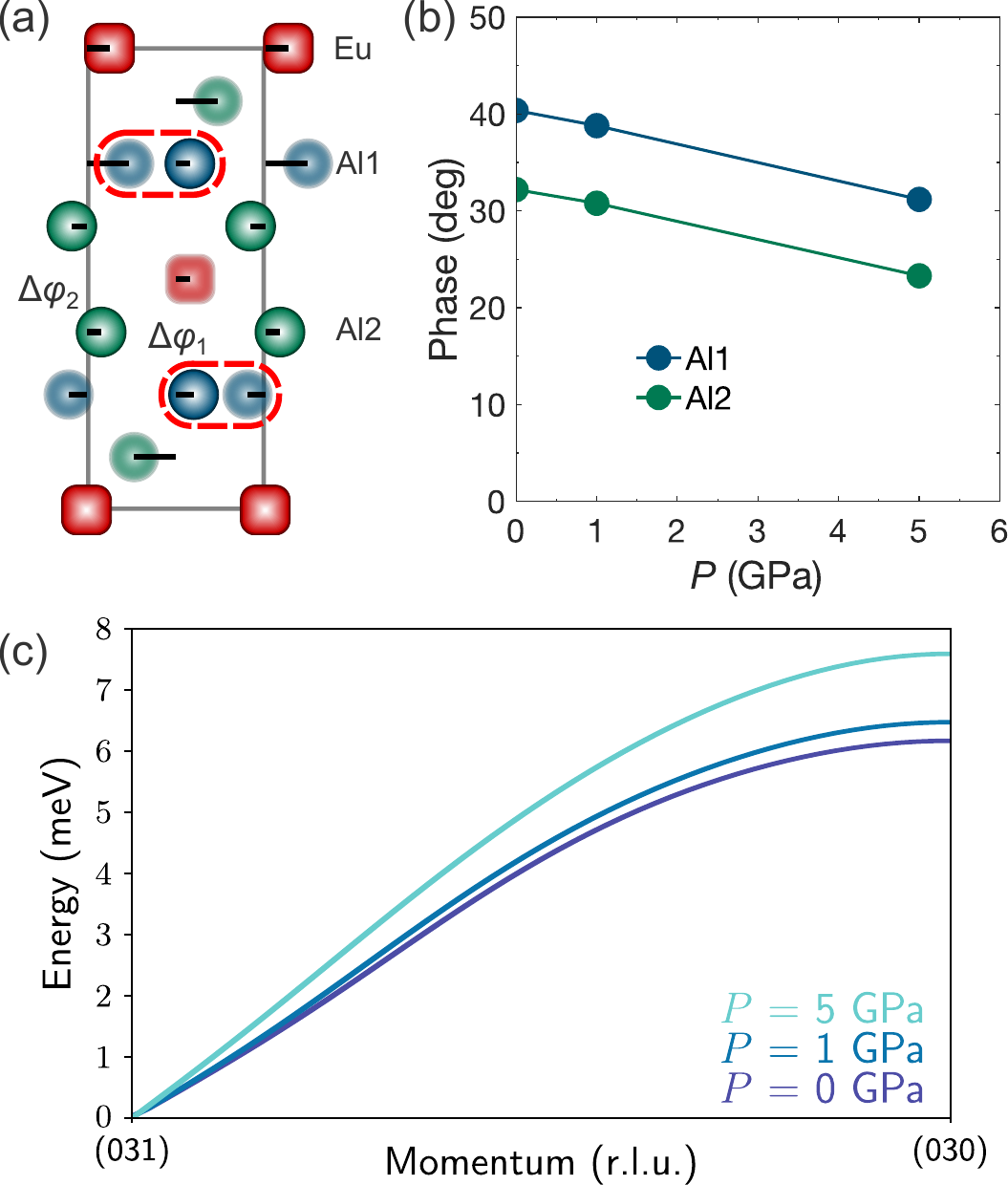}}
  \caption{The results of the \textit{ab initio} calculations. (a) The sketch of the atomic displacements of the TA phonon at the momentum (0~0~0.2) r.l.u. The red dotted lines show which Al atoms form the zigzag chains. (b) The pressure-induced change of the relative atomic phases in the mode. (c) The calculated TA mode dispersion under different pressures.}
  \label{fig:Fig4}
  \vspace{-12pt}

\end{figure}

The lattice dynamics of EuAl$_4$ under hydrostatic pressure were calculated using the projector-augmented wave method~\cite{PhysRevB.59.1758} and density functional theory as implemented in the \textsc{vasp} software~\cite{PhysRevB.54.11169,KRESSE199615} similarly to the calculations at the ambient pressure reported earlier~\cite{PhysRevB.110.045102}. We used the generalized gradient approximation functional with Perdew-Burke-Ernzerhof parametrization~\cite{PhysRevLett.77.3865}, a $8\times 8\times 8$ $k$-point mesh (Monkhorst-Pack scheme~\cite{PhysRevB.13.5188}) for Brillouin zone integration, and a plane-wave cutoff of 400~eV. Phonon band dispersions were calculated with finite difference method using \textsc{phonopy}\cite{TOGO20151} on a $(3\times 3\times 4)$ supercell of the primitive unit cell. 

In the simulations under high pressure, we focus on the transverse acoustic mode measured in the experiment. In the previous study~\cite{PhysRevB.110.045102}, the displacements of the mode in the vicinity of the CDW wavevector were shown to resemble the structure of the CDW. Figure~\ref{fig:Fig4}(a) depicts such atomic displacements within the first unit cell (the full wavelength of the mode is five unit cells). All the atomic positions within the primitive cell, which contains one Eu and four Al atoms, can be parameterized by only two parameters: the phase offset on the Al1 sublattice ($\Delta\phi_1$) and the phase offset on the Al2 sublattice ($\Delta\phi_2$). For convenience, the phase on the first Eu atom (left bottom corner) was set to zero. The phase offset of the two remaining Al atoms is opposite to $\Delta\phi_1$ and $\Delta\phi_2$, respectively. The atomic positions within the next primitive cells is then determined by the respective translation vectors.

As can be seen in Fig.~\ref{fig:Fig4}(a), the phase delay on the Al1 sublattice causes formation of atomic zigzag-chains with shortened Al1--Al1 distances within the $ab$ plane. These zigzag-chains have varying stretch along the propagation vector; if a phonon is polarized along the $a$ axis, the zigzag-chains of Al1 atoms run along $b$. In Fig.~\ref{fig:Fig4}(b), we traced the displacement vectors at the same momentum under pressure and plotted the $\Delta\phi_1$ and $\Delta\phi_2$ phases. Evidently, the phase delays on both the Al1 and Al2 sublattices diminish under pressure. In effect, the mode transforms into a ``more primitive'' transverse wave where the atoms of each layer are being displaced in phase. This implies that the Al1-Al1 bonding is relaxed under pressure, which agrees with the experimental observation of CDW suppression. Even though the calculated phase change is quite small for pressures at which the experimental EPC shows suppression, it provides qualitative agreement with the experimental data.

Figure~\ref{fig:Fig4}(c) shows the renormalization of the mode dispersion under $P = 1$ and 5~GPa. The first-principle calculations without EPC predict very little hardening $\sim$5\%/GPa of the mode due to the trivial change of the force constants under the unit cell compression. This further supports the conclusion that the experimental spectrum renormalization is dominated by the EPC rather than by an ordinary pressure influence on the (uncoupled) lattice dynamics.

\section{Discussion}

This IXS study revealed a pronounced suppression of the mode-softening in EuAl$_4$ under pressure. Evidently, the reduction of EPC in this system depends on both temperature and pressure. Thus, the basic trends of its variation can be traced within the $(T$--$P)$ phase diagram. As  discussed above, the EPC strength can be estimated via Eqs.~\ref{eq:eq1} and \ref{eq:eq2}, where the amplitude parameter $A$ is determined via fits to the experimental data. We can combine the values of $A$ obtained under high pressure with the values determined from the previous measurements at ambient pressure~\cite{PhysRevB.110.045102}. Moreover, because the maximal (limiting) electron-phonon coupling strength at which the mode freezes out, is known, we can attribute this value, $A_{\text{max}} = 2.6$\,meV to each phase transition $(T$--$P)$ point reported by~Nakamura \textit{et al.}~\cite{Nakamura:2015aa}.

\begin{figure}[t]
\center{\includegraphics[width=1.0\linewidth]{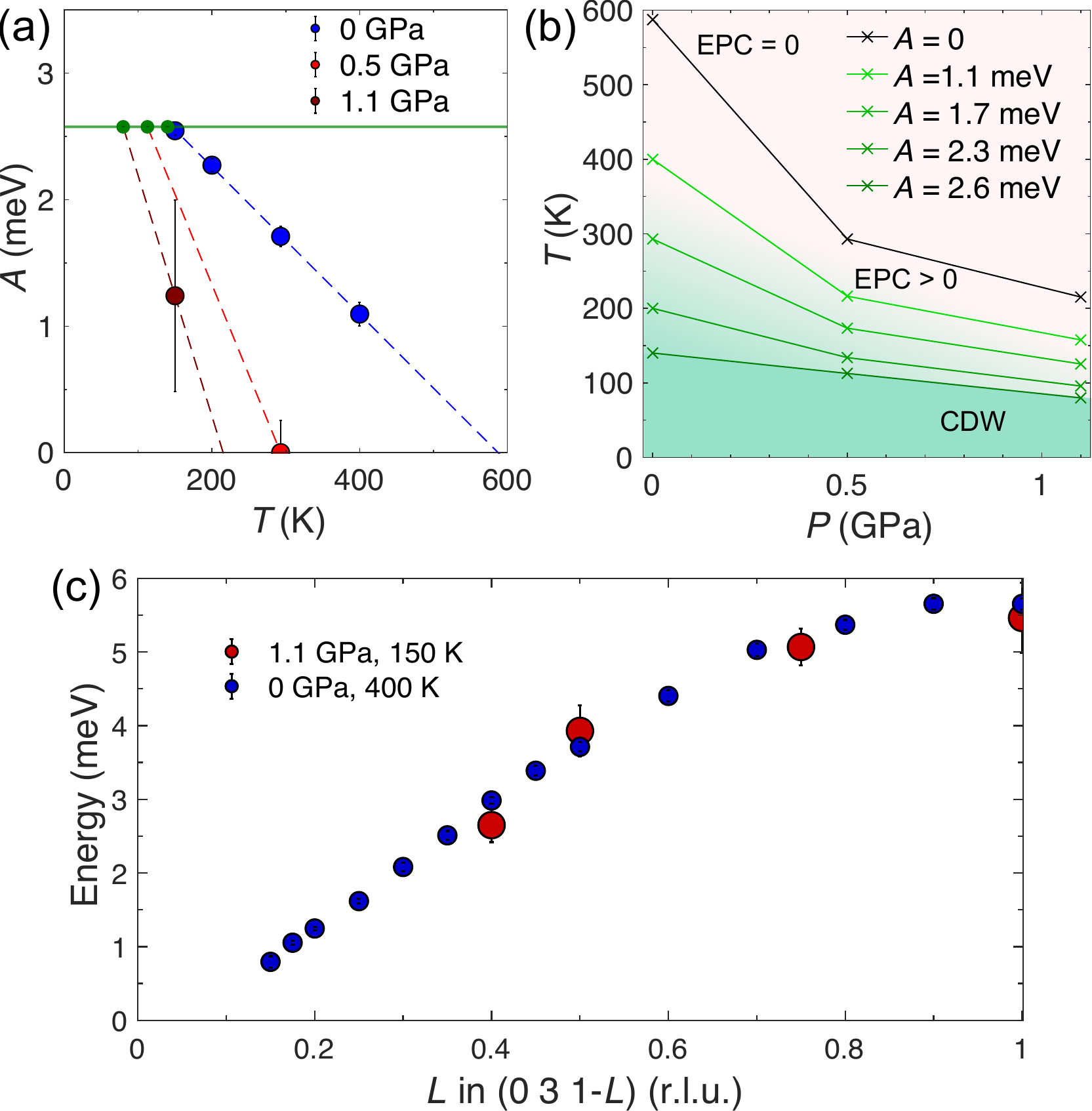}}
  \caption{(a) Temperature dependence of the scale parameter $A$ of Eq.~\ref{eq:eq2}, proportional to the EPC strength, for different pressures. For each $P$, the green symbols mark the values $A_{\text{max}}$, at which the softened mode freezes out, and the green solid line is a guide to the eye indicating this limit. (b) The corresponding contours of constant $A$ within the $(T$--$P)$ phase diagram of EuAl$_4$. (c) The similarity of the phonon dispersion measured under different $T$ and $P$ conditions, as suggested by (b).}
  \label{fig:Fig5}
  \vspace{-12pt}

\end{figure}

In this fashion, Fig.~\ref{fig:Fig5}(a) summarizes the thermal variation of the EPC amplitude $A$ at different pressures. Since $A(T)$ displays a clear linear trend at ambient pressure, we speculate that a similar linear trend is preserved under pressure as well, as indicated by the dashed lines. Following this assumption, one can draw some conclusions on iso-EPC contours in the phase diagram of EuAl$_4$, as illustrated in Fig.~\ref{fig:Fig5}(b). At a glance, this shows how the EPC is renormalized under pressure; specifically, with increasing pressure, its thermal decline is compressed into a narrower margin of lower temperatures. Furthermore, this ``compression'' is nonlinear, it proceeds faster at smaller pressures, and then slows towards higher $P$. It follows from this tentative analysis of Fig.~\ref{fig:Fig5}(b) that the EPC strength, and thus the phonon softening, at $P = 1.1$~GPa and $T = 150$~K should be approximately equal to that of 400~K at ambient pressure. Remarkably, the data collected at these two conditions indeed overlap exactly, as shown in Fig.~\ref{fig:Fig5}(c).

It is worth comparing EuAl$_4$ with other known CDW materials where the phonon softening was investigated under high pressure. For example, the room-temperature measurements on elemental uranium showed that the softening of the optic $\Sigma_4$ mode along the reciprocal ($h$00) direction remains quite pronounced until pressure as high as 20~GPa, whereas the CDW is fully suppressed already at $\sim$1.5~GPa~\cite{PhysRevLett.107.136401}. This can be strongly contrasted to EuAl$_4$, where both the mode softening and the CDW order suppressed on very similar pressure scale. A somewhat similar pressure-induced behavior was observed in the transition-metal dichalcogenide 2$H$-NbSe$_2$~\cite{PhysRevB.92.140303}, where the LA mode softening remained up to pressures of 16~GPa, as compared to the critical pressure of the CDW suppression at 4.6~GPa. It is important to note that the extended mode softening in 2$H$-NbSe$_2$ was attributed to the anharmonicity effects~\cite{PhysRevB.92.140303}, which may suggest that EuAl$_4$, in contrary, can be well treated in the harmonic approximation. 

\section{Conclusion}

To summarize, we observe a systematic suppression of the CDW transition in EuAl$_4$ with increasing pressure, accompanied by a significant reduction in the EPC strength, which confirms the transport-measurements of~\cite{Nakamura:2015aa}. Inelastic X-ray scattering reveals that even at moderate pressures of 0.5~GPa, the phonon softening is notably diminished, and that this suppression becomes more pronounced at $\sim$1~GPa. First-principles simulations further support these findings, showing that pressure weakens the characteristic atomic displacements associated with the CDW state, in particular, by reducing the preferential bonding between Al atoms in the Al1 layers. Importantly, the pressure-induced changes in phonon dispersion are predominantly driven by the renormalization of EPC rather than by a trivial stiffening of phonon modes due to the unit cell compression.

\section*{Acknowledgements}

We thank A.~Bosak for help with sample preparation and planning of IXS experiments. The experimental work was performed at end-station ID28 of the European Synchrotron Radiation Facility (ESRF). Work at TUD was supported by the German Research Foundation (DFG) through the CRC1143 (project-id 247310070) and the W\"{u}rzburg--Dresden Cluster of Excellence ct.qmat (EXC2147, project-id 390858490). MCR is grateful for support through the Emmy-Noether programme of the DFG (project-id 501391385). A.K. acknowledges the Education Department of the Basque Government (Grant No. PIBA\_2023\_1\_0051). SG is grateful for support through the graduation scholarship program of the Association of Friends and Sponsors of TU Dresden. Work at Rice University has been supported by the Robert~A.~Welch Foundation under grant no.~C-2114.

\clearpage

\bibliographystyle{apsrev4-1} 
\bibliography{EuGaAl}
\end{document}